# Recommendation Scheme Based on Converging Properties for Contents Broadcasting

Jian Sun
Department of Electronic Engineering
Tsinghua University
Beijing, CHINA
sunj13@mails.tsinghua.edu.cn

Xuan Zhou
Department of Information Science & Electronic Engineering
Zhejiang University
Hangzhou, CHINA
zhouxuan@zju.edu.cn

Xiaofeng Zhong
Tsinghua National Laboratory for Information Science and Technology
Department of Electronic Engineering, Tsinghua University
Beijing, CHINA
zhongxf@tsinghua.edu.cn

Xiaolong Fu
Information Technology Center
Tsinghua University
Beijing, CHINA
fxl@tsinghua.edu.cn

***Abstract*—Popular videos are often clicked by a mount of users in a short period, and take most of the network traffic. With content recommendation, the popular contents could be broadcast to the potential users in wireless network, to save huge transmitting resource, as described in Content Aware Soft Real Time Media Broadcast (CASoRT) system. With the users' communication log in the Chinese commercial cellular network, the contents propagation model is analyzed due to users' historical behavior, location, and the converging properties in wireless data transmission is shown in this paper. Then, a recommendation scheme is proposed to achieve high energy efficiency in CASoRT system.**

***Keywords—content propagation character; recommendation scheme; converging property; CASoRT; broadcast; energy efficiency***

## I. Introduction

Multiple-media data service has long been a key application in wireless cellular network, which always requires higher data rate and consumes more energy and bandwidth. But the existing wireless cellular network is only designed to try its best to transmit bit, rather than content. In fact, the users' behavior of media requirement is highly related to converging property in web contents, users' interest and users' location. For example, there may be plenty of users who browse the same web page of one hot news in a period of time, and the wireless cellular network needs to transmit the same data of the hot news many times to many different subscribers, which will lead to many redundancy transmitting in same cell, in very low efficiency way.

Based on this observation, [1] proposes the Content Aware Soft Real Time Media Broadcast (CASoRT) system, which broadcasts selected contents to proper groups of users, to save much more wireless transmitting bandwidth. Additionally, [7] proposes the broadcast scheme for CASoRT system which achieves suboptimal energy efficiency. Obviously, an important issue of CASoRT is the recommender system, which figures out the real-time pre-push strategy according to the web access logs of mass users, and pushes appropriate web contents to appropriate range of users. Thus, when the covered users need to browse the pushed web contents, which is likely to happen under the recommendation algorithm, they can acquire them from their local memory instead of another transmission in wireless cellular network, and the network traffic could be alleviated.

The efficiency of the recommender system will decide the performance of CASoRT. But unfortunately, traditional recommender systems focus on personalization rather than group behavior of mass users. With the increase of user numbers and visits, their performance will be degraded. Therefore, statistical analysis of mobile users' behavior is necessary, and a new efficient recommendation scheme for CASoRT exploiting the converging properties in wireless data transmission should be proposed. In this paper, the network logs are gathered from Chinese commercial cellular communication system, which covers more than 75 million users. The content propagation model is analyzed, and a content recommendation scheme is proposed, due to users' historical behavior and users' location, which could be implemented in the CASoRT system.

This paper is organized as follows. Section II presents the related work. Section III introduces network log data gathering and processing procedures. Section IV presents the converging properties in wireless data transmission, from the perspective of users, web contents, and geographic positions. Section V proposes the recommendation scheme, and shows the network performance improvement implemented in CASoRT system. Section VI is the conclusion.

## II. Related Work

Although the roots of recommender systems can be traced back to the extensive work in cognitive science, approximation theory, information retrieval, forecasting theories, and also have links to management science and consumer choice modeling in marketing, recommender systems emerged as an independent research area in the mid-1990s [2]. Basically, the strategy of recommendation falls into three categories: collaborative filtering recommendation, content-based recommendation, and hybrid recommendation. The collaborative filtering algorithm was based on the point that people who share similar interest in the past would still share interest in the future [3], while the content-based algorithm will recommend items similar to the ones the user preferred to in the past, and hybrid recommendation combine the above two methods [2]. A series of techniques and applications based on these recommendation schemes has been developed over the past few years, and performed well in many fields. [8] proposes a K neighbors collaborative filtering prediction method, which improves the efficiency of traffic prediction by 28% comparing with basic collaborative filtering.

Users' behavioral characteristics do be important to recommendation scheme. [4] proposes a technique that uses user behavior monitoring to transparently capture the users' interest in information, and a technique to use this interest to filter incoming information in a very efficient way. [5] proposes the Online Behavioral Analysis and Modeling Methodology to accurately and efficiently categorize users based on their individual web browsing activities. [6] analysis the access logs accumulated by a unique web service, and get some insight into the behavior of mobile users regarding the spatial information on the web. However, the behavioral analysis of mass mobile users in wireless cellular network, from the perspective of users, web contents, and geographic locations, is still lacking.

## III. Log Data Processing

### *A. Data Gathering*

The users' data access logs are gathered from commercial wireless cellular network of China Mobile Communication Corporation (CMCC), which is the largest mobile operator in China, as well as in the world. The access logs is orientated from more than 35 million users of Zhejiang Province and contains User ID in amorous code, start-stop time of data session, URL, Location Area Code and Cell ID for location information. The web visits of 7 consecutive days, from April 19 to April 25, 2014, is illustrated in figure 1.

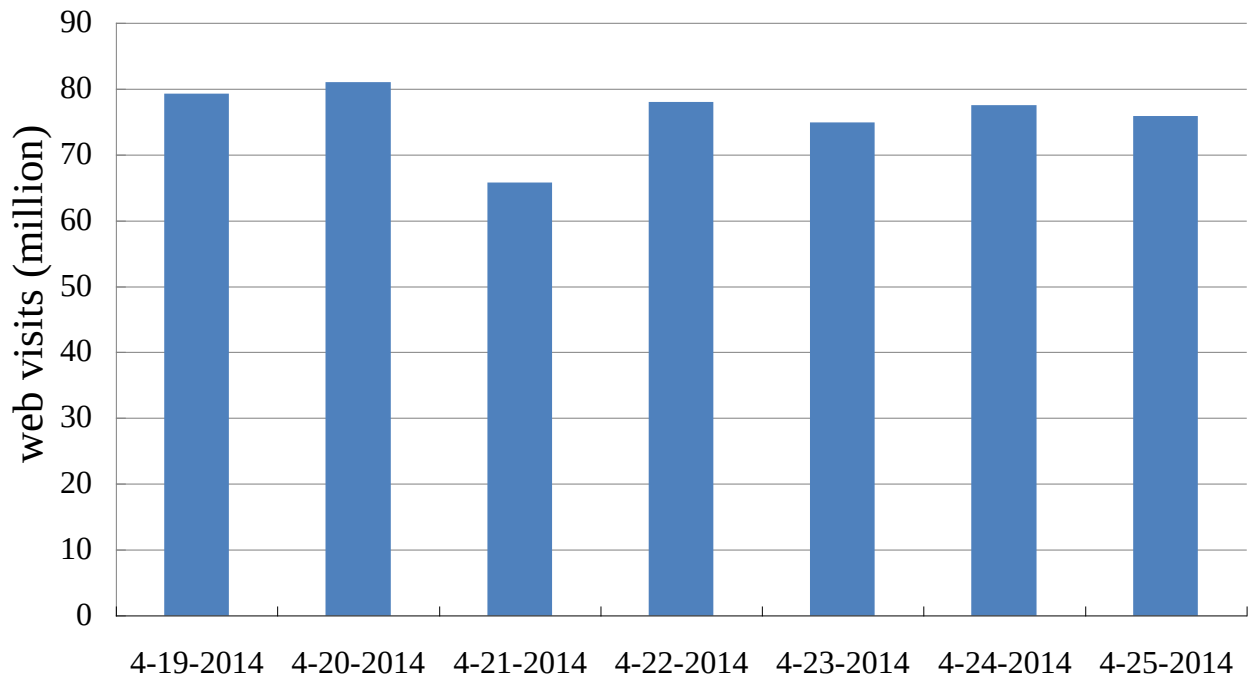

Fig. 1. Web visits of commercial cellular users

The web visits of these days is rather enormous, nearly 80 million visits per day in average. Therefore, to match our data processing ability, there will be a data filtration before the statistical analysis. According to different contents, we choose several websites for news, novels, and comprehensive information under the following three principles. 1. The accessibility of web contents. There may be some protected web contents which can't be accessed by a third party via URL addresses, and also outdated URLs are useless. We filter out the access logs with such URLs in the first place, due to the web contents browsed by users hold significant value to our analysis. 2. The representativeness of web contents. We choose the web contents which browsed by a great number of mainstream users, and avoid those mainly browsed by niche users, for the purpose of guaranteeing the universality of our research. For example, news websites provide suitable web contents for our research. 3. On the basis of the previous principles, the data volume should be limited to a reasonable range. Oversize data will be difficult to manipulate, while too little data wouldn't be representative.

The websites we chose include China's fourth largest internet portal owned by Phoenix New Media. With the domain name of "ifeng.com", this portal is a sound, video, images and comprehensive information website. Later analysis will be based on the access logs of this chosen website, whose visits within 7 days is illustrated in figure 2.

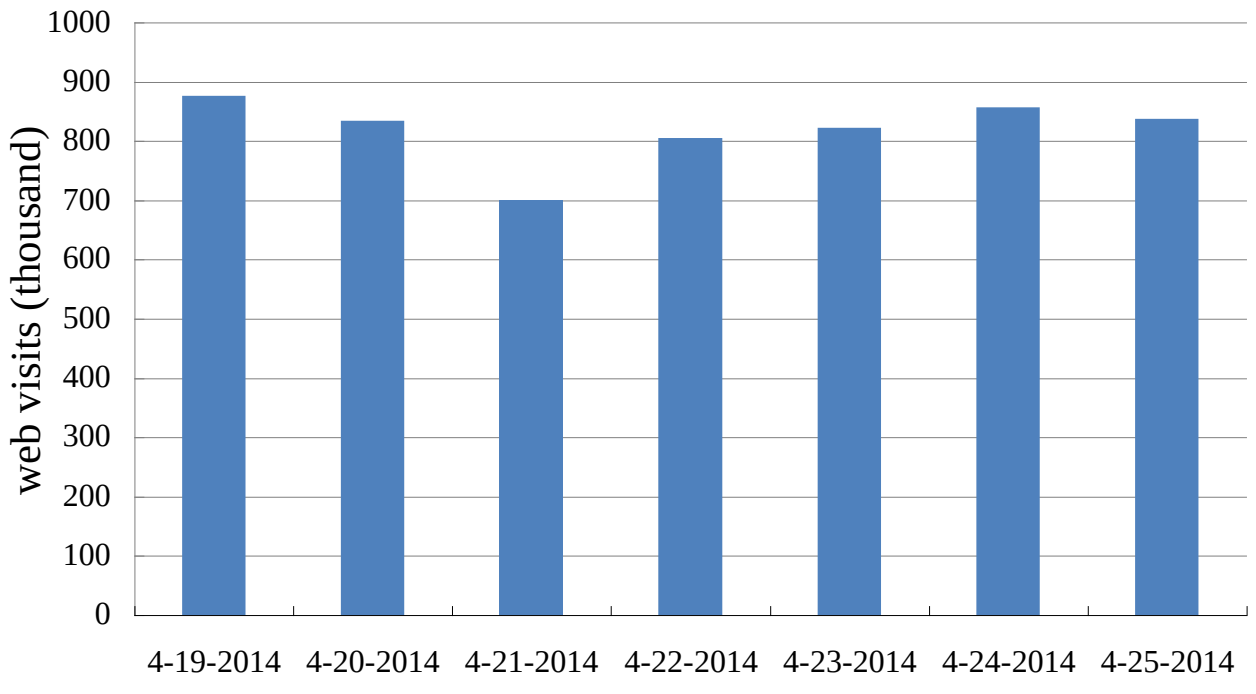

Fig. 2. Web visits to "ifeng" website

The network traffic of ifeng is about 800 thousand accesses per day in average, which is suitable for our analysis. Verified by URLs random selected from the access logs of ifeng, we confirmed that the vast majority of its web contents is reachable. What's more, the rich content and wide range of user groups of ifeng ensures its representativeness. These factors mentioned above make ifeng an appropriate choice.

### *B. Data Processing*

The purpose of data processing is to get the web contents browsed by users, which are just represented by URLs in access logs. In this process, we use RabbitMQ (a kind of message queue) to distribute data processing tasks, and then use a web crawler to access the web contents via URLs. We finally get 3,462,339 complete web access logs of ifeng within 7 days, stored in MongoDB (database).

There are two steps in data processing: obtain the HTML files by the web crawler, and extract the main content of the HTML files by some web content information extraction method. Several synergetic computers are used to process over 5 million access logs of ifeng within an acceptable time.

In data processing, one of those computers acts as a host, deploying with the producer program which constantly sends complete access logs one by one to the message queue. The others act as clients, deploying with the consumer program, which receives the access logs of mobile users and process them in the following way. After receiving a web access log, the consumer program downloads the HTML file via the URL in the log, and extracts the main content (a title and a main body in general) of the HTML file by the extraction method. And then, it stores the main content to the file system which will return a unique file identifier. Thus, all the fields needed for analysis of this very access log are acquired, and the last step of data processing is to store the objects to MongoDB.

## IV. CONVERGING PROPERTY ANALYSIS

### *A. Converging Property of Users*

It is an intuitive idea that there are active users and inactive users among all the mobile users, and the active users are the key concern naturally. In this section, we will carry on the quantitative analysis about the influence of active users on web visits. Extracted from the 3,462,339 access logs of the ifeng, we have 262,055 distinct users identified by IMSI. By sorting the users by activity level, and accumulating the web visits from the most active users, we can get the relationship between the most active users and the percentage of web visits they consume, as illustrated in figure 3.

According to calculation, the top 10% of the most active users consume more than 50% of the web visits, and the top 20% of the most active users consume nearly 70% of the web visits. This means that it is the minority of the active users who consume most of the network traffic.

Therefore, it will be rather cost-effective for the recommender system to focus limited computing resources and broadcasting energy on a minority of the most active users in a period of time.

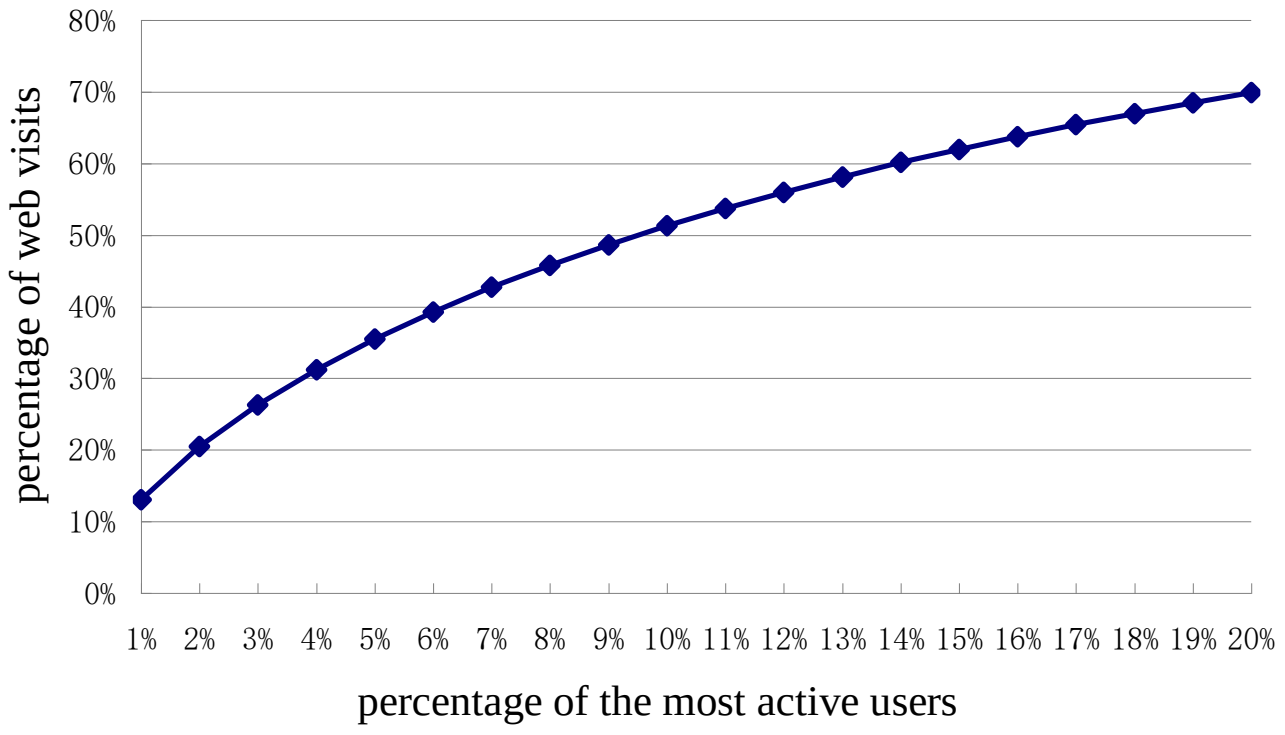

Fig. 3. The converging property of users

### *B. Converging Property of Web Contents*

Based on the results of the previous section, we can expect the similar converging property when we inspect the web contents browsed by users. The web page titles obtained during the data processing, rather than the URLs which often lead to the same web page, should be used to identify the different web contents. We get 36,804 distinct titles from the post-processing data, and the relationship between the most popular titles and the percentage of web visits they consume is illustrated below.

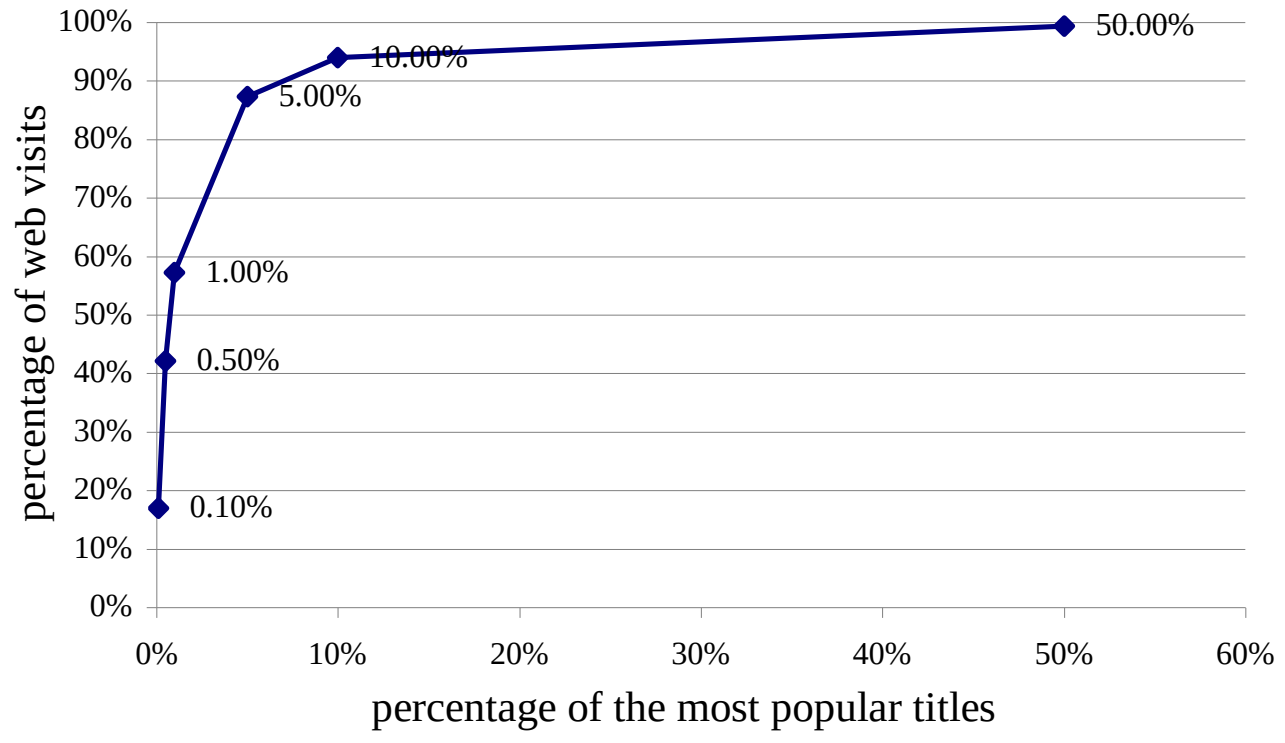

Fig. 4. The converging property of web contents

As is shown in figure 4, the top 0.1% of the most popular titles (37 titles) consume about 17% of the web visits, and with the rapid rise of the polyline, the top 5% of the most popular titles consume nearly 90% of the web visits. In other words, a small collection of the most popular titles has a significant impact on network traffic. It's obvious that the converging property of web contents is much stronger than that of users. This could be explained by the influence of the hot news, which attract a large number of visits.

Therefore, the recommender system should focus on the popular web contents to alleviate the network traffic with fewer extra resources. Due to the converging properties of users and web contents, if the recommendation algorithm is aware of the active users and the popular contents in real time, the size of user-traffic matrix will be dramatically reduced, and make it possible for the recommender algorithm to meet the real-time requirements of CASoRT system.

## C. *Converging Property of Geographic Positions*

Geographic position has been a key concern in cellular network recently. In this section, we will focus on the converging property of geographic positions in wireless data transmission, and figure out its value for our recommendation scheme. We obtain the converging property of geographic positions (LACs and CIDs), as illustrated in figure 5.

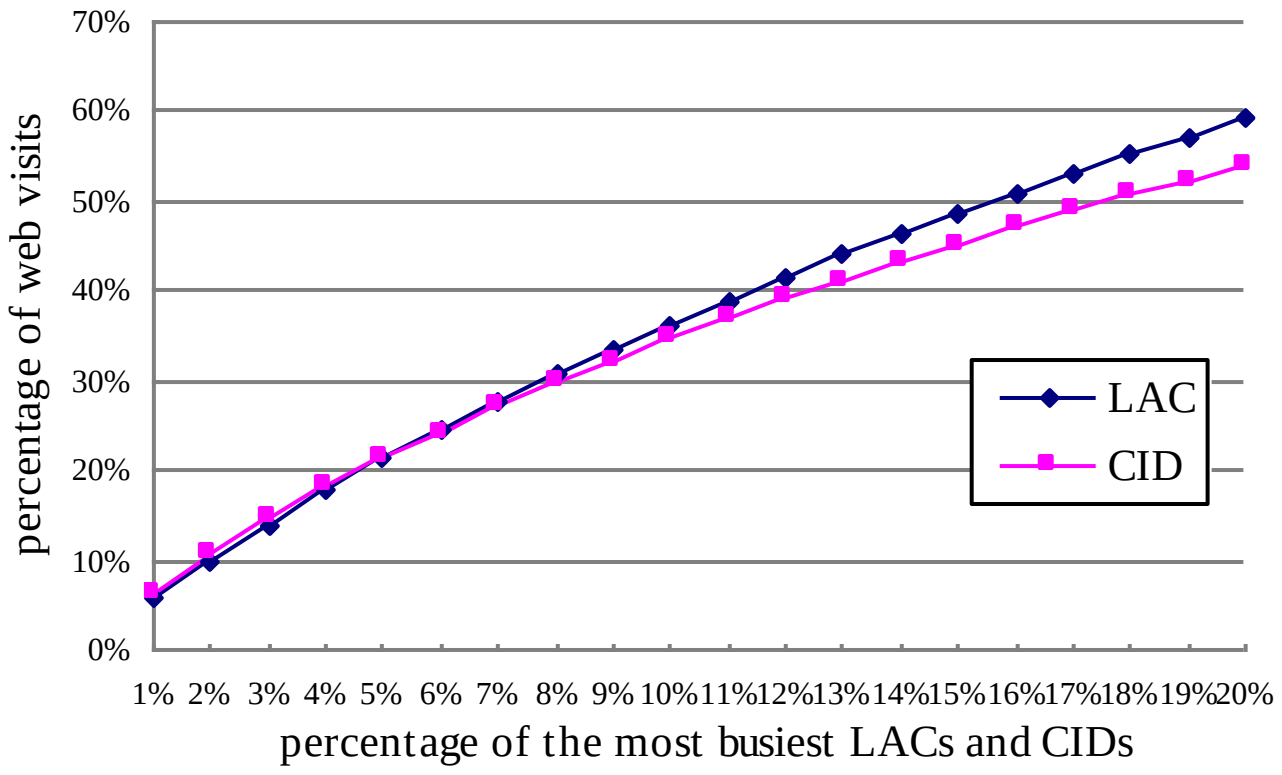

Fig. 5. The converging property of geographic positions

Compared with the analysis results of users and web contents, the converging property of geographic positions is not evident, which means focusing the computing resources and broadcasting energy of the recommender system on some busy areas only brings limited improvement. But if we pay close attention to single mobile user, analyze its geographic distribution, the converging property can still be observed clearly. We sample 500 active users randomly and count the active cells of each one, and the statistical result is shown in figure 6.

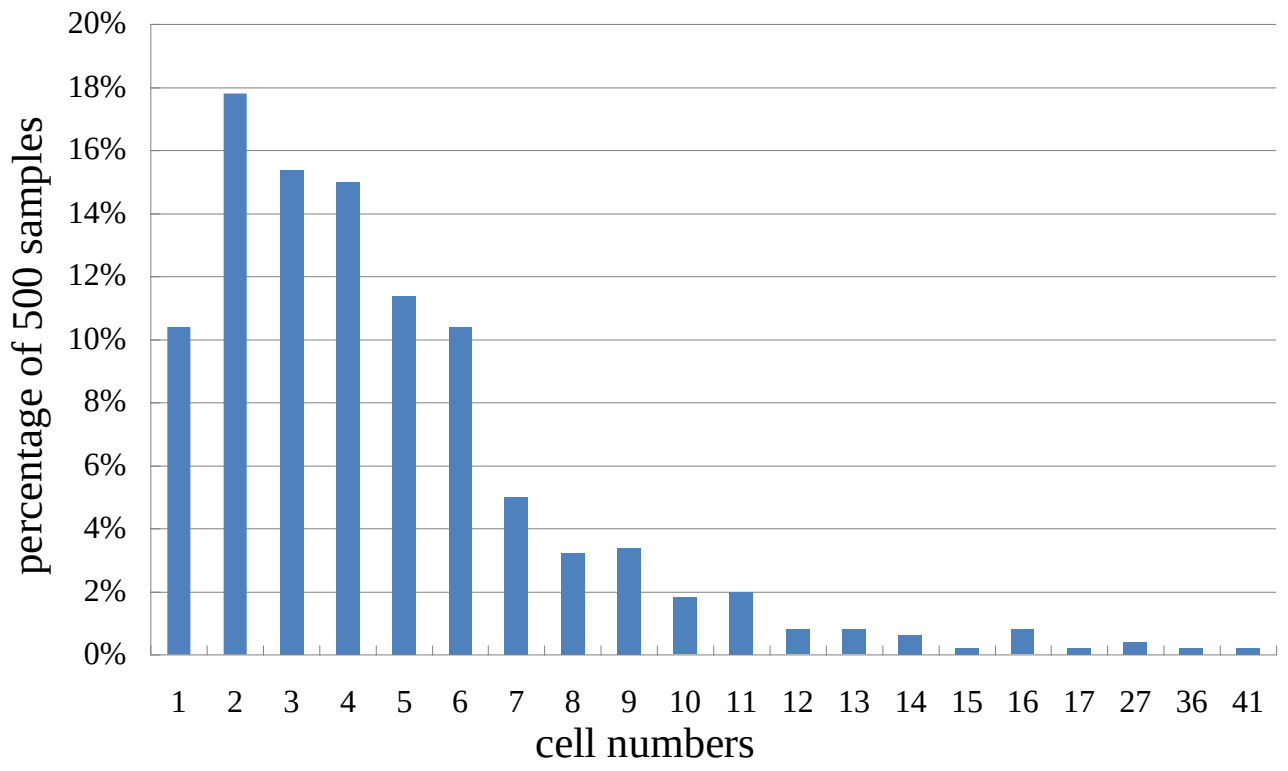

Fig. 6. The distribution of active cell numbers

The calculation shows that one user has 4.8 active cells in average, and most users have less than 10 active cells. By observing the geographic distribution of each user, we find that in average, user's most active cell bears about 58% of its web visits, and top 2 of the most active cells bear about 80%, as is shown in table I. This geographic converging property of single user is rather important while pushing data to user's local memory, which will be further discussed in the next section.

TABLE I. GEOGRAPHIC CONVERGING PROPERTY OF SIGNGLE USER

| Number Of Cells | Percentage Of Web Visits In Average |
|---|---|
| 1 | 58% |
| 2 | 80% |
| 3 | 89% |
| 4 | 94% |
| 5 | 96% |

# V. RECOMMENDATION SCHEME BASED ON CONVERGING PROPERTIES

In this section, we evaluate the recommendation scheme based on the converging properties above, and show the network performance improvement in CASoRT system by numerical analysis.

While pushing data to user's local memory, the system has to "guess" which cell the user is in. Since active cells of single user are limited, and 1 or 2 cells bear most of the network traffic, according to table I, it is reasonable to assume that the user is just in its most active cell, though the consequence of this assumption should be investigated.

The CASoRT system performance could be investigated in 3 cases. In case 1, both coverage and accuracy of the recommendation algorithm are 100%, and all the users' actual positions are knowable to our system. This means that the recommendation algorithm is able to predict the visits of any title in our data, and the system is able to push data to any user as needed, both accurately. In case 2, the recommendation algorithm remains the same, but the system just assume that all the users are in their most active cells. This means that not all the target users will receive the data sent to them. In case 3, all the conditions are the same as case 2, except the coverage of the recommendation algorithm is 20%, which means that for any title in our data, 20% of its visitors could be predicted accurately. And we assume that they are the top 20% of the most active users.

## A. *Case 1*

To begin with, we analyze the most popular title in our data. This title has 30,749 visits, thus the data of this web page has been transmitted 30,749 times to different users. And in fact there are 17,217 distinct users distributed throughout 13,833 distinct cells who browsed this web page, which means that the CASoRT system will be able to broadcast the data to all users by 13,833 data transmissions (each broadcast in a cell counts as one transmission), instead of 30,749 times independently, according to the assumptions. Thus the number of data transmissions will be reduced by 16,916.

There are 36,804 distinct titles in our data, and the more titles are broadcasted, the more reduction of data transmissions could be achieved. According to our observation, popular titles will contribute most of the reduction of data transmissions, because the redundancy in some cells is evident for these titles.

For example, if the system broadcasts the top 1%, namely 368 titles to users who will visit them, the number of data transmissions will be reduced to 1,062,956. Plus the visits from the other 99% titles, the total number of data transmissions will be 2,543,859, 73% of the original visits. Figure 7 shows the relationship between the network traffic and the popular titles broadcasting ratio.

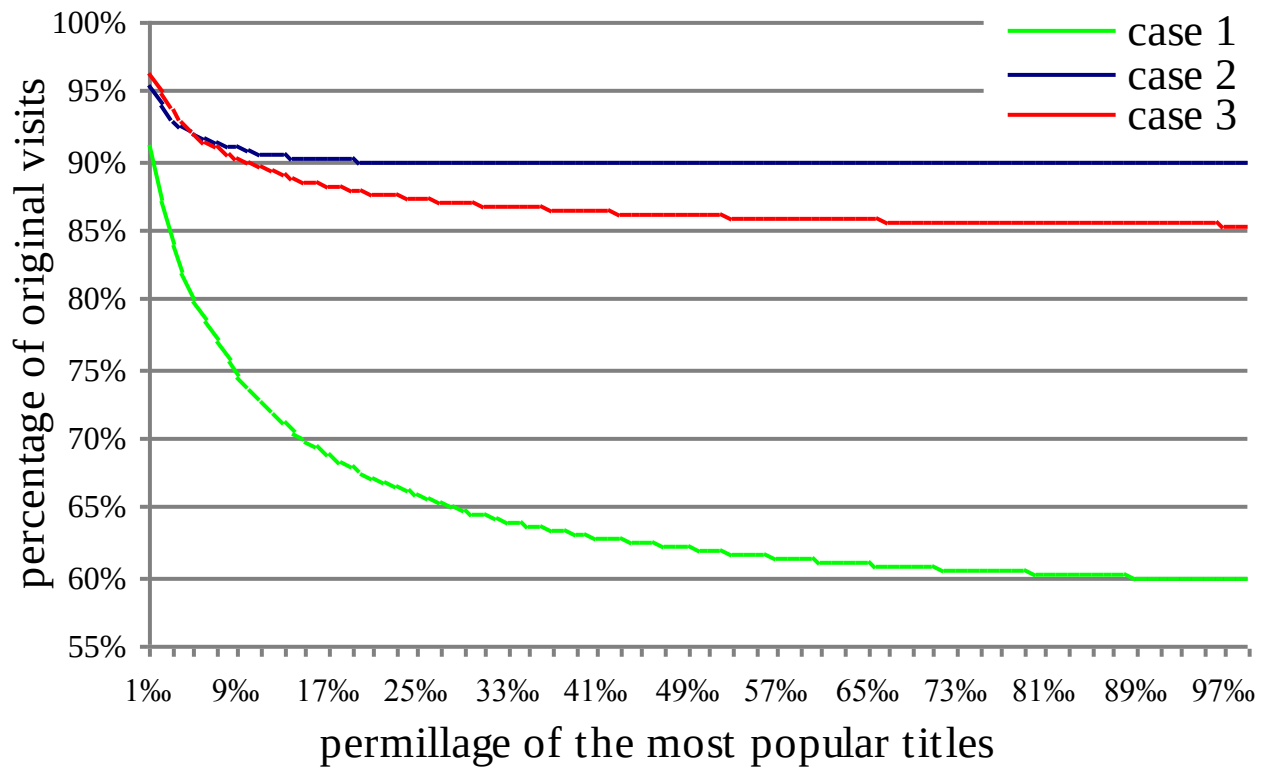

Fig. 7. The performance of CASoRT system in 3 cases

As shown in figure 7, the network traffic decreases rapidly with more popular titles broadcasted, and becomes steady while broadcasting less popular titles. Generally, the reduction of network traffic is obvious under these circumstances.

## B. *Case 2*

We analyze the most popular title with 17,217 visitors first. By recording each user's most active cell, we have a collection of 13,489 cells in which the data of this web page are meant to be broadcasted, and thus 13,489 data transmissions are needed for broadcasting. Compared with 13,833 cells in which the data transmissions are actually happened, as shown in figure 8, we find that 2,999 cells are mistaken, 3,343 cells are missing, and the visits in these missing cells is 5,795. Thus, the number of data transmissions for this web page will be reduced to 19,284.

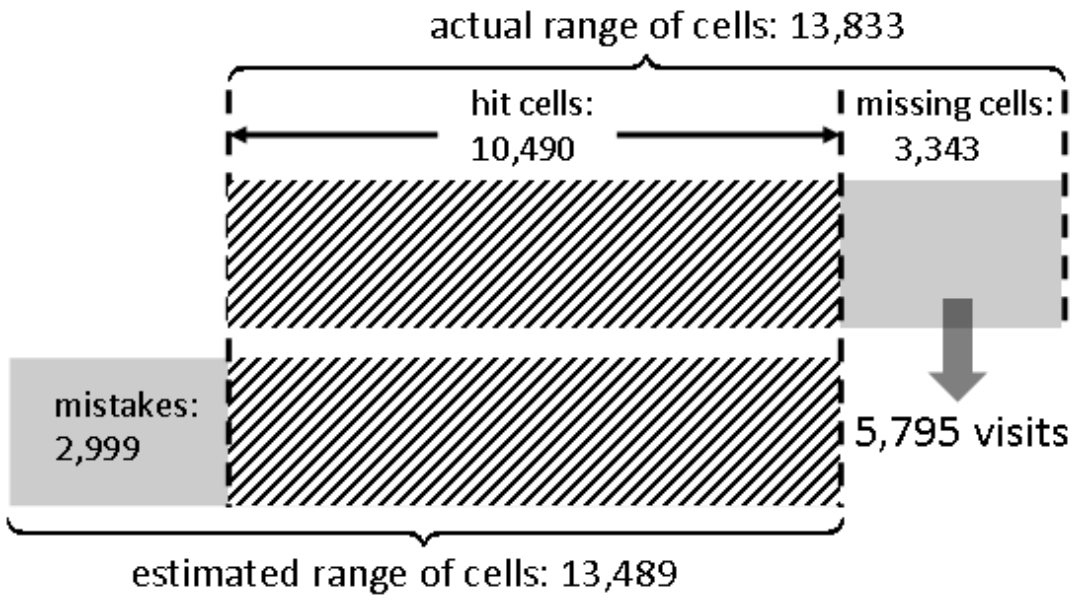

Fig. 8. The hit cells and missing cells in estimation

As in case 1, we calculate the reduction of network traffic while broadcasting more titles, and the result is shown in figure 7 (blue line). Unexpectedly, we find that the network traffic remains nearly constant while broadcasting less popular titles to all the users who will visit them, and the reduction of network traffic is not evident. This result will be explained in the next section.

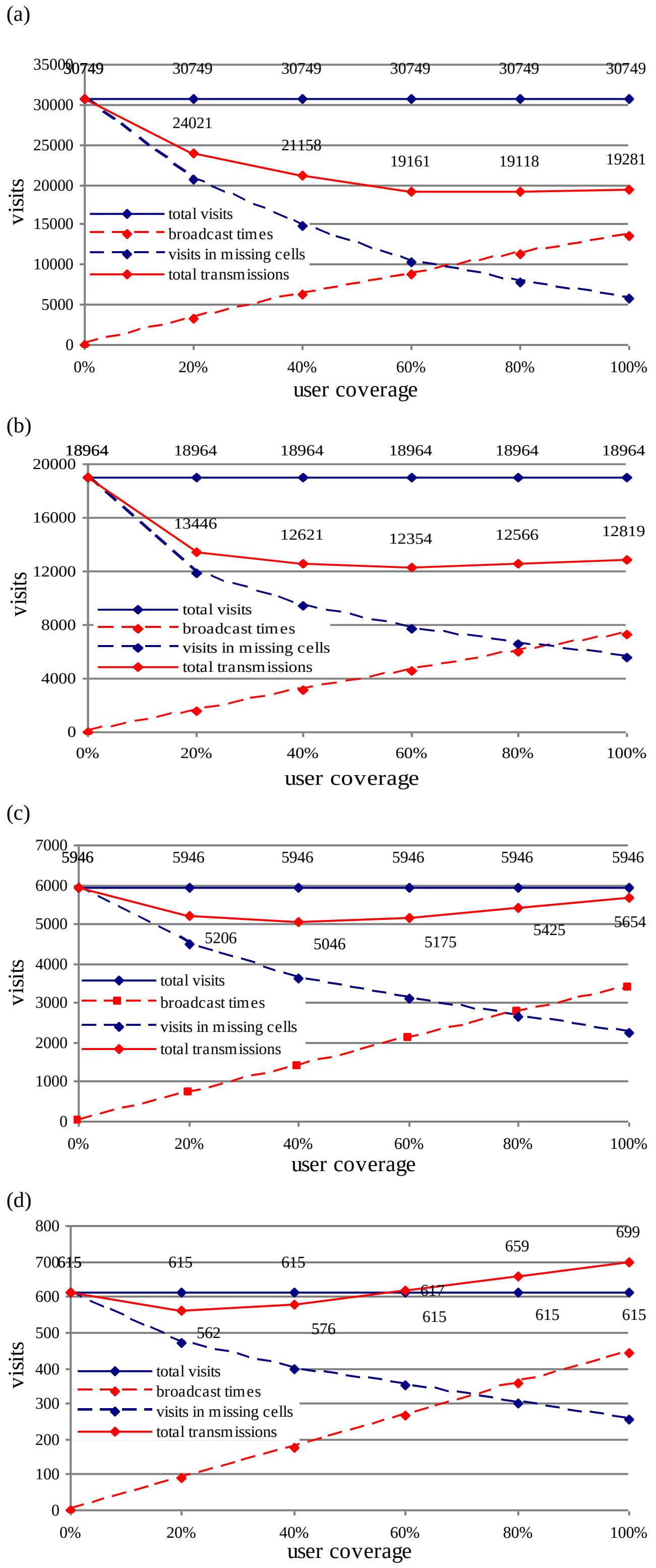

Fig. 9. The reduction of data transmissions with different user coverage. (a) The most popular title. (b) 10th most popular title. (c) 100th most popular title. (d) 1000th most popular title.

### C. *Case 3*

In this case, the coverage of the recommendation algorithm is 20%, thus the calculation method is the same as case 2 except the broadcast range is limited to the top 20% users. The system performance is better than that in case 2, according to the calculation result shown in figure 7 (red line). The explanation of this counter-intuitive phenomenon is that while broadcasting less popular titles with less redundancy, the bigger the estimated range of cells, the less the missing cells, but the more mistakes at the same time, thus the reduction of data transmissions will be counteracted.

The blue lines in Figure 9 indicate the original visits of each title (4 typical titles are illustrated). The red lines, which are the sum of the blue dashed lines and the red dashed lines, indicate the data transmissions needed for each title with the CASoRT system works under different user coverage. The red dashed lines indicate the data transmissions needed for broadcast, and the blue dashed lines indicate the visits in the missing cells, accordingly. With the increase of user coverage, the CASoRT system needs to do more broadcasts, and the missing cells decrease.

As shown in figure 9, the optimal user coverage is about 80% for the most popular titles (the lowest point of the red line in figure 9a), and 60%, 40%, 20% for the other three titles (figure 9b, 9c, 9d), respectively. So, the data of popular titles should be broadcasted to relatively more potential users, while the broadcast range of less popular titles should be restrained, otherwise the CASoRT system will be counterproductive, as shown in figure 9d.

Therefore, it is neither realistic nor cost-effective for CASoRT system to cover all the users, or broadcast all web contents. Overall, limited broadcast range will be conductive to the performance of CASoRT system.

## VI. CONCLUSION

This paper presents the converging properties in wireless data transmission, from the perspective of users, web contents, and geographic positions. According to the statistical results, we draw the conclusion that the convergence is evident in users, web contents, and single user's geographic distribution. Based on these properties, an efficient recommendation scheme is proposed for CASoRT system, and the numerical results show that the alleviation of network traffic is obvious while broadcasting appropriate web contents to appropriate range of users.

The CASoRT system should focus on popular web contents and active users to achieve the reduction of network traffic, and avoid less popular web contents and inactive users which will lead to low system performance. This strategy is consistent with the Pareto principle, and also reduces the requirements of the recommendation algorithm. What's more, if the CASoRT system is aware of users' positions, the reduction of network traffic is remarkable. And if it is not, it's a simple and effective way to assume that the cellular user is just in its most active cell while pushing data to its local memory.

## ACKNOWLEDGMENT

This work is supported by National Basic Research Program of China (2012CB31600), and National S&T Major Project (2013 ZX03003004-002).